\documentclass[twocolumn,showpacs,prb,aps,superscriptaddress]{revtex4}
\usepackage{amsmath}
\usepackage{graphicx}
\usepackage{color}
\newcommand{\be}{\begin{equation}}
\newcommand{\ee}{\end{equation}}

\begin{document}
\title{Complex magnetic ordering in CeGe$_{1.76}$ studied by neutron diffraction}
\author{W. T. Jayasekara}
\affiliation {Ames Laboratory, USDOE and Department of Physics and Astronomy, Iowa State University, Ames, Iowa 50011, USA}
\author{W. Tian}
\affiliation{Quantum Condensed Matter Division, Oak Ridge National Laboratory, Oak Ridge, Tennessee 37831, USA}
\author{H. Hodovanets}
\affiliation {Ames Laboratory, USDOE and Department of Physics and Astronomy, Iowa State University, Ames, Iowa 50011, USA}
\author{P. C. Canfield}
\affiliation {Ames Laboratory, USDOE and Department of Physics and Astronomy, Iowa State University, Ames, Iowa 50011, USA}
\author{S. L. Bud'ko}
\affiliation{Ames Laboratory, USDOE and Department of Physics and Astronomy, Iowa State University, Ames, Iowa 50011, USA}
\author{A. Kreyssig}
\affiliation{Ames Laboratory, USDOE and Department of Physics and Astronomy, Iowa State University, Ames, Iowa 50011, USA}
\author{A. I. Goldman}
\affiliation {Ames Laboratory, USDOE and Department of Physics and Astronomy, Iowa State University, Ames, Iowa 50011, USA}

\date{\today}

\begin{abstract}

Neutron diffraction measurements on a single crystal of CeGe$_{1.76}$ reveal a complex series of magnetic transitions at low temperature. At $T_{\rm{N}}$ $\approx$ 7~K, there is a transition from a paramagnetic state at higher temperature to an incommensurate magnetic structure characterized by a magnetic propagation vector (0~0~$\tau$) with $\tau$ $\approx$ $\frac{1}{4}$ and the magnetic moment along the \emph{\textbf{a}} axis of the orthorhombic unit cell.  Below $T_{\rm{LI}}$ $\approx$ 5~K, the magnetic structure locks in to a commensurate structure with $\tau$ = $\frac{1}{4}$ and the magnetic moment remains along the \emph{\textbf{a}} axis.  Below $T^{\star}$ $\approx$ 4~K, we find additional half-integer and integer indexed magnetic Bragg peaks consistent with a second commensurately ordered antiferromagnetic state.

\end{abstract}

\pacs {}

\maketitle

\section{\label{Intro} INTRODUCTION}

The physics of competing interactions and their effect on the ground state of magnetic systems has remained a strong focus of interest in the scientific community. The CeGe$_{2-x}$ compounds are prototypical systems that lie close to an instability between ferromagnetic (FM) and antiferromagnetic (AFM) ordering. This is indicated by the near zero value of the Weiss temperature, $\theta$, and the proximity of the Ce-Ce distance to the critical value for the boundary between FM and AFM order \cite{Lin_2002}. Furthermore, the Sommerfeld coefficient evaluated in CeGe$_{2-x}$ ($\gamma >$ 100 mJ/mol K$^{2}$) is significantly enhanced and qualifies CeGe$_{2-x}$ as a moderate heavy fermion system.  For all of these reasons, CeGe$_{2-x}$ has been the subject of many structural and magnetic investigations over the past five decades.\cite{Matthias_1958,Gladyshevskii_1959,Gladyshevskii_1964,Eremenko_1971,Eremenko_1972,Yashima_1982,Mori_1985,Gokhale_1989,Schobinger_1991,Lambert_1994,Venturini_1999,Lin_2002,Zan_2003,Shcherban_2009,Zhang_2013,Budko_2014}

The $R$Ge$_{2-x}$ compounds form over a range of Ge concentrations ($x$ $\approx$ 0 to 0.4). Samples with $x$ $>$ 0.3 crystallize in the tetragonal ThSi$_2$-type structure (space group $I4_1/amd$) whereas samples with higher Ge content ($x$ $<$ 0.3) crystallize in the orthorhombic GdSi$_2$-type structure (space group $Imma$) shown in Fig.~\ref{Fig1}. However, a recent comprehensive survey \cite{Zhang_2013} of $R$Ge$_{2-x}$ compounds illustrated how partial or full ordering of the Ge-site vacancies gives rise to a variety of superstructures.  In particular for $x$ = $\frac{1}{4}$, the  NdGe$_{2-x}$\cite{Venturini_1999}, PrGe$_{2-x}$\cite{Shcherban_2009} and SmGe$_{2-x}$\cite{Zhang_2013} compounds form in the vacancy-ordered $R_4$Ge$_7$ structure with (\textbf{\emph{ab}}) plane dimensions four times larger than the parent orthorhombic cell.  For $R$ = La and Ce, on the other hand, no long-range ordering of the vacancies was found, although selected-area electron diffraction images showed evidence of regions of both commensurate and incommensurate modulations in the (\textbf{\emph{ab}}) plane of the orthorhombic structure.\cite{Zhang_2013}

The range of superstructures and, in some cases, the coexistence between the orthorhombic and two distinct tetragonal structures within the same sample\cite{Lambert_1994} has led to confusion and conflicting results concerning the structure and physical properties of this class of compounds, For polycrystalline CeGe$_{2-x}$ samples with $0 \leq x \leq$ 0.29 ($Imma$), an AFM transition with $T_{\rm{N}} \approx$ 7~K followed by a FM transition with $T_{\rm{C}} \approx$ 4~K was reported.\cite{Zan_2003} For lower Ge concentrations, $0.29 \leq x \leq$ 0.34 ($I4_1/amd$), only a single AFM transition with $T_{\rm{N}} \approx$ 7~K was found.\cite{Zan_2003}  These results, however, were contradicted by later studies,\cite{Nakano_2005} where two magnetic transitions, AFM ordering at 6.7~K followed by a FM transition at 5.3~K, were found for tetragonal CeGe$_{1.66}$.  Recently, an extensive study of the physical properties of single crystals of CeGe$_{1.76}$, including resistivity, heat capacity, dc susceptibility, thermal expansion, and thermoelectric power has been reported,\cite{Budko_2014} which found evidence of three, rather than two transitions at $\approx$ 7~K, $\approx$ 5~K and $\approx$ 4~K, respectively.

Here, we describe the results of neutron diffraction measurements of the magnetic ordering in CeGe$_{1.76}$ single crystals produced by the solution-growth technique\cite{Canfield_1992} which identify three magnetic transitions in this system. At $T_{\rm{N}}$ $\approx$ 7~K, there is a transition from a paramagnetic state to an incommensurate magnetic structure characterized by a magnetic propagation vector (0~0~$\tau$) with $\tau$ $\approx$ $\frac{1}{4}$ and the magnetic moment along the orthorhombic \emph{\textbf{a}} axis. Upon further cooling below $T_{\rm{LI}}$ $\approx$ 5~K, the magnetic structure locks in to a commensurate structure with $\tau$ = $\frac{1}{4}$ and the magnetic moment remains along the \emph{\textbf{a}} axis.  Below $T^{\star}$ $\approx$ 4~K, where FM ordering has previously been suggested by magnetization measurements\cite{Zan_2003,Budko_2014}, we find additional half-integer and integer indexed magnetic peaks, consistent with a second commensurately ordered AFM state.  However, given the small value of the ferromagnetically ordered moment of $\approx$0.2 $\mu_{\rm{B}}$/Ce reported in Ref.~\onlinecite{Budko_2014}, we were unable to verify the presence of a FM component.

\section{\label{ExpDetails} Experimental Details}

\begin{figure}
\centering\includegraphics[width=0.6\linewidth]{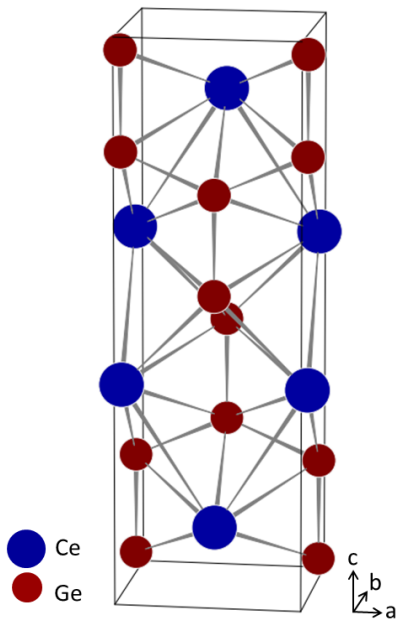}
\caption{(Color online) The body-centered orthorhombic unit cell of CeGe$_{2}$ (space group $Imma$).  The blue spheres represent Ce and the maroon spheres Ge.}
\label{Fig1}
\end{figure}

Single crystals of CeGe$_{1.76}$ were grown using a high-temperature solution technique \cite{Canfield_1992} as described in detail elsewhere \cite{Budko_2014}. The crystals grow as large plates, with dimensions in excess of $7~\times~7~\times~1$~mm$^3$ with the \textbf{\emph{c}}-axis perpendicular to the plate.  Elemental analysis was performed on these crystals using wavelength-dispersive x-ray spectroscopy (WDS) in the electron probe microanalyzer of a JEOL JXA-8200 electron microprobe. Ce$_2$Fe$_{17}$ and elemental Ge were used as the standards. The stoichiometry of the samples was determined to be CeGe$_{1.76(1)}$. Room temperature powder x-ray diffraction measurements were taken on samples from the same batch as used for the neutron diffraction measurements and the study described in Ref.~\onlinecite{Budko_2014}. The diffraction pattern were recorded employing Cu-radiation in a Rigaku Miniflex diffractometer and have been refined using the Rietica software.\cite{Hunter_1998} The x-ray data are consistent with an orthorhombic unit cell (space group $Imma$) with the lattice parameters $a$ = 4.338(9) \AA, $b$ = 4.248(8) \AA, and c = 14.04(3) \AA, similar to values previously reported.\cite{Gladyshevskii_1964,Mori_1985,Schobinger_1991,Venturini_1999} For reference, the conventional orthorhombic unit cell of CeGe$_{2}$ is shown in Fig.~\ref{Fig1}.

Single-crystal neutron diffraction measurements were performed on a 175 mg single crystal of CeGe$_{1.76}$ using the HB-1A triple-axis spectrometer at the High Flux Isotope Reactor at Oak Ridge National Laboratory. The experiment
was performed using a fixed incident neutron energy of 14.6 meV with collimations of 40$^{\prime}$-40$^{\prime}$-40$^{\prime}$-80$^{\prime}$ before the pyrolytic graphite monochromator, between the monochromator and sample, between the sample and pyrolytic graphite analyzer, and between the analyzer and detector, respectively. Two pyrolytic graphite filters were used to eliminate higher harmonics in the incident beam. The single crystal was mounted in a helium cryostat (1.5 - 300 K) with the (0~$K$~$L$) reciprocal lattice plane coincident with the scattering plane to investigate the magnetic structure and with the ($H$~$K$~0) plane coincident with the scattering plane to search for chemical superlattice reflections characteristic of the $R_4$Ge$_7$ structure.  We note that the presence of structural domains and twinning in the orthorhombic structure\cite{Budko_2014} means that scans taken nominally along the in-plane [0~$K$~0] reciprocal lattice direction also capture scattering along the [$H$~0~0] direction of the other structural domain.  We, therefore, adopt the labeling convention of [0~$K$~0] ([$H$~0~0]$^{\prime}$) for scans of this type.

\section {Results}

\begin{figure*}
\centering\includegraphics[width=1.0\linewidth]{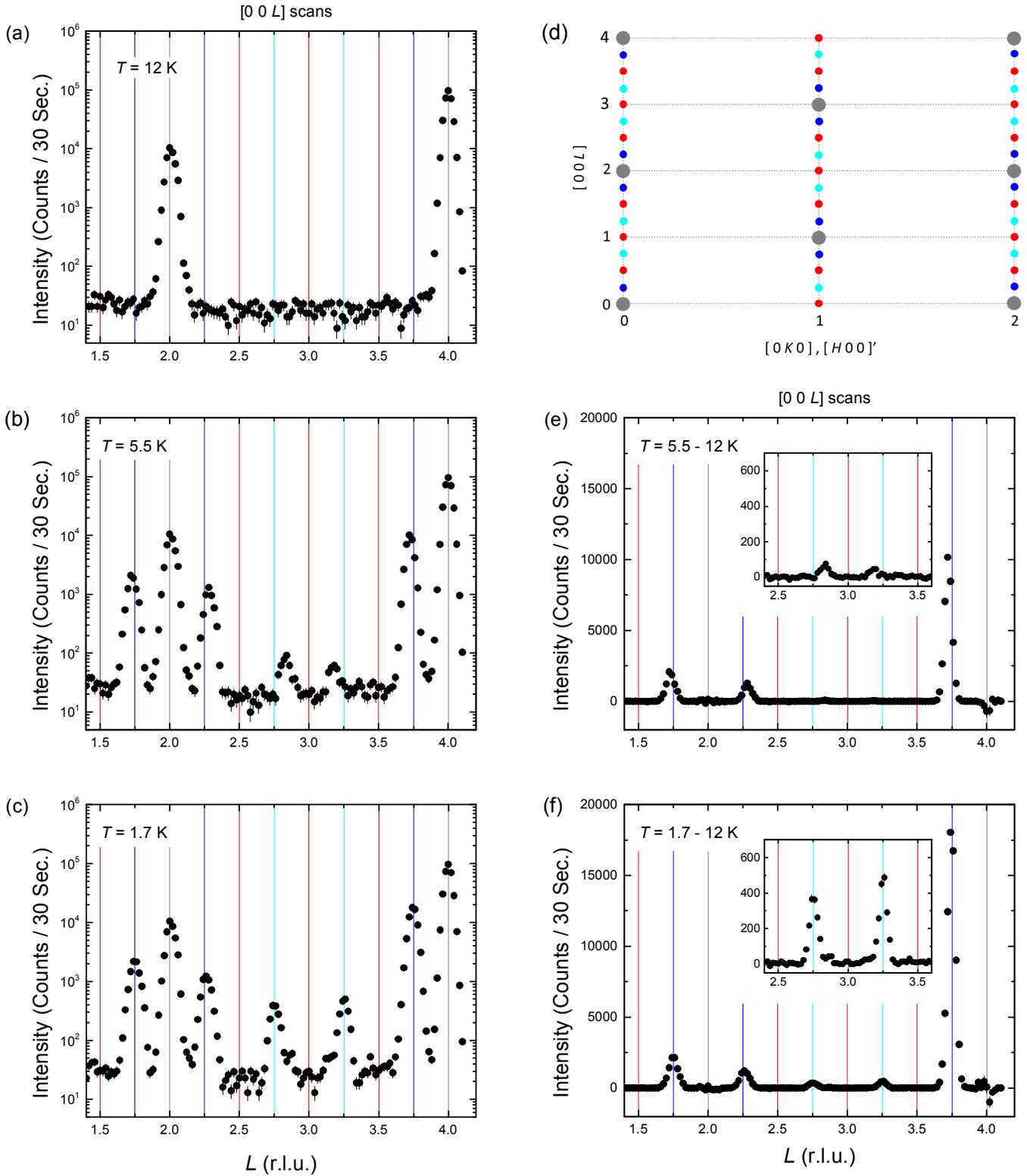}
\caption{(Color online) Neutron diffraction measurements on CeGe$_{1.76}$. Panels (a), (b) and (c) show scans along the [0 0~$L$] reciprocal lattice direction on a logarithmic scale at sample temperatures of 12~K (above $T_{\rm{N}}$), 5.5~K (below $T_{\rm{N}}$ but above $T_{\rm{LI}}$) and 1.9~K (below $T_{\rm{LI}}$), respectively, illustrating the evolution of the (0~0~$\tau$) and (0~0~3$\tau$) magnetic Bragg peaks. A map of the (0~$K$~$L$) reciprocal lattice plane is shown in panel (d). Filled gray symbols represent the allowed nuclear Bragg peaks. Solid dark blue symbols represent the magnetic Bragg peaks observed near, and at, the magnetic wave vector (0~0~$\tau$) with $\tau$ $\approx$ $\frac{1}{4}$ for $T<T_{\rm{N}}$. The light blue solid symbols show the positions of higher harmonic (0~0~3$\tau$) magnetic Bragg peaks. The solid red symbols represent additional magnetic Bragg peaks at half-integer and integer positions (not observed or very weak in the [0~0~$L$] scans). The difference, on a linear scale, between the data taken at 5.5~K and at 12~K; and 1.9~K and 12~K are shown in panels (e) and (f). The insets to panels (e) and (f) highlight the region near the (0~0~3$\tau$) magnetic Bragg peaks. The vertical lines denote the positions of half-integer and integer peaks in addition to the $\tau$ = $\frac{1}{4}$ and $\frac{3}{4}$ commensurate peak positions, as shown in panel (d).}
\label{Fig2}
\end{figure*}

\begin{figure}
\centering\includegraphics[width=0.8\linewidth]{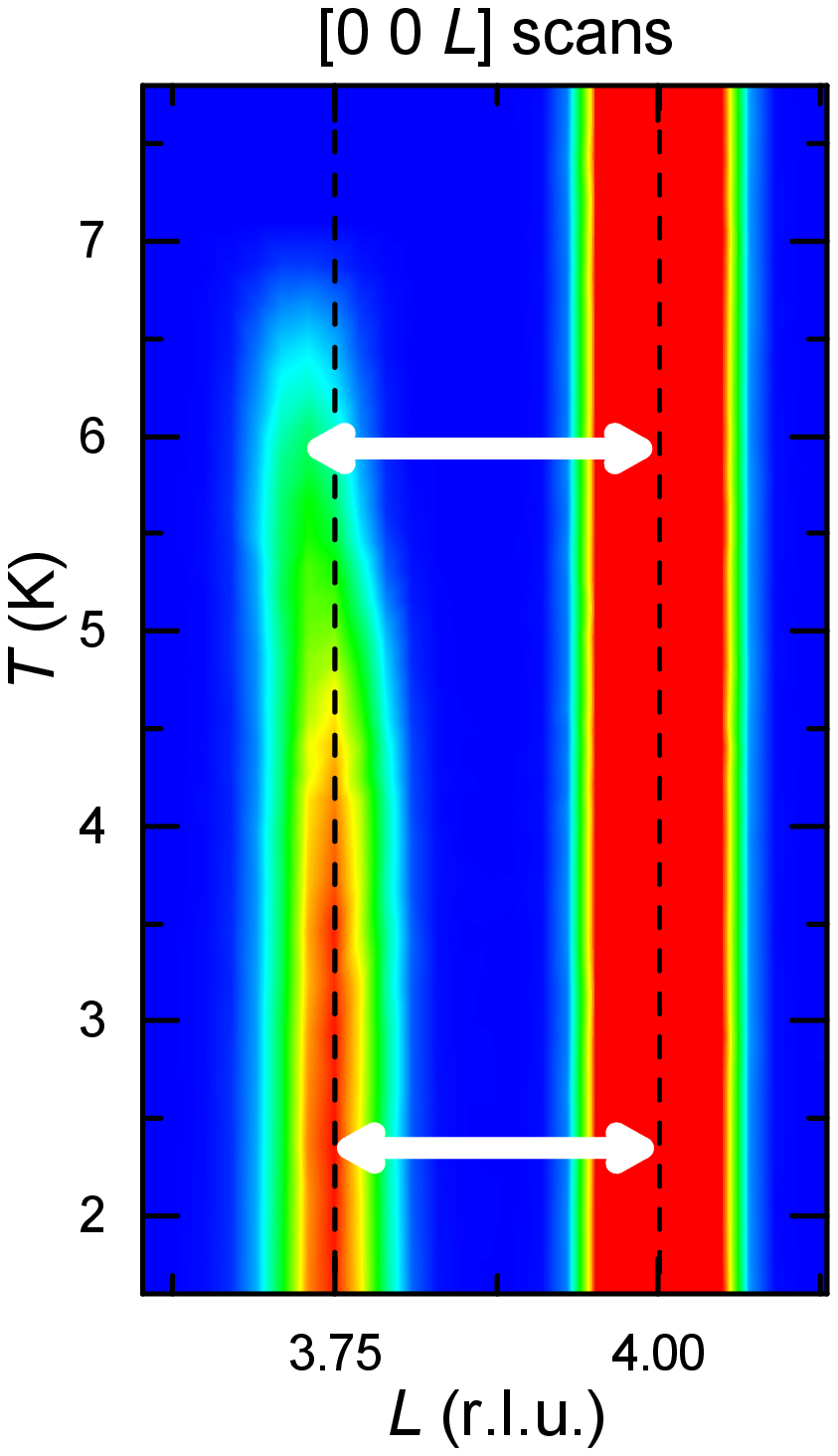}
\caption{(Color online) Contour plot of the (0 0 4-$\tau$) magnetic Bragg peak as a function of temperature above and below the lock-in transition at $T_{\rm{LI}}$ $\approx$ 5~K. The arrows illustrate how $\tau$ changes above and below $T_{\rm{LI}}$.}
\label{Fig3}
\end{figure}

\begin{figure}
\centering\includegraphics[width=0.9\linewidth]{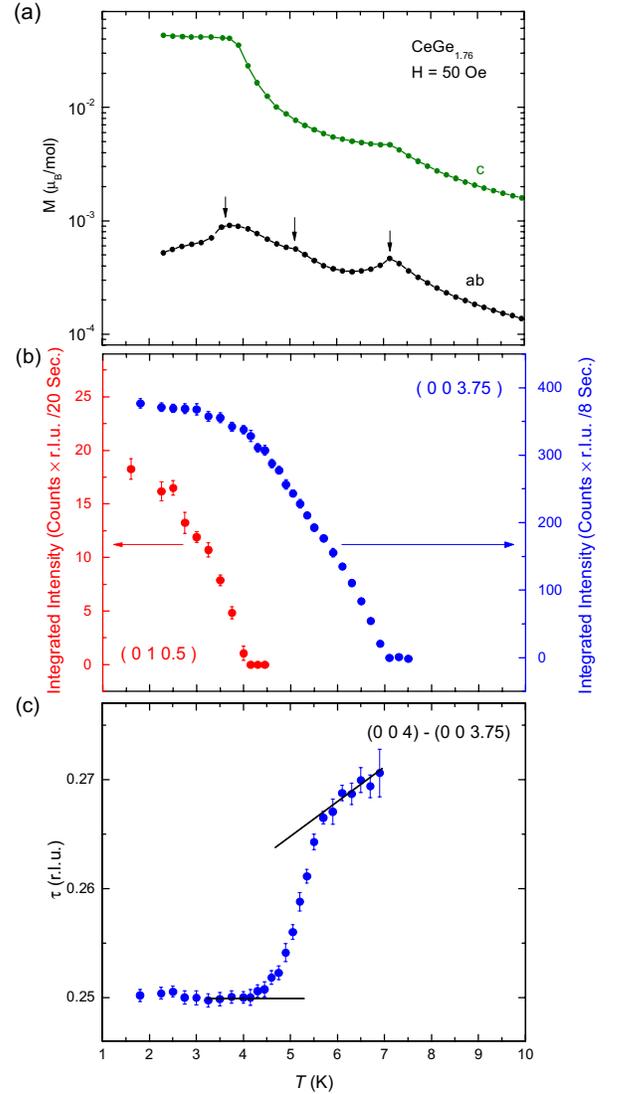}
\caption{(Color online) The temperature evolution of (a) the bulk magnetization (after Ref.~\onlinecite{Budko_2014}), (b) the integrated intensities of the (0~0~4-$\tau$) and (0~1~0.5) magnetic Bragg peak, (c) the value of $\tau$ measured  for the (0~0~4-$\tau$)  magnetic Bragg peak. The solid lines serve as guides to the eye.}
\label{Fig4}
\end{figure}

For temperatures ($T$) above $T_{\rm{N}}$, Fig.~\ref{Fig2}(a) shows that only nuclear reflections [gray solid symbols in Fig.~\ref{Fig2}(d)] consistent with the body-centered orthorhombic reflection condition, $H$ + $K$ + $L$ = 2$n$ ($n$ = integer), were observed in the (0~$K$~$L$) reciprocal lattice plane.  Upon cooling the sample below $T_{\rm{N}}$, but above $\approx$ 5~K, Fig.~\ref{Fig2}(b) and Fig.~\ref{Fig2}(e) demonstrate that additional magnetic reflections [solid dark blue circles in Fig.~\ref{Fig2}(d)] appear at positions corresponding to a magnetic propagation vector (0~0~$\tau$) with $\tau$ $\approx$ $\frac{1}{4}$.  At the same time, a third harmonic (0~0~3$\tau$) satellite appears and grows in intensity [solid light blue symbols in Fig.~\ref{Fig2}(d)], indicating that the incommensurate structure is "squaring up" as $T$ is lowered.  Figures~~\ref{Fig2}(c) and (f), as well as Figs.~\ref{Fig3} and \ref{Fig4}(c), show the evolution of $\tau$ as a function of $T$.  Upon lowering $T$ below $T_{\rm{N}}$, $\tau$ first slowly decreases from its initial value of 0.27, followed by a steep decrease between 4.5~K and 5.5~K to a value of $\frac{1}{4}$. The midpoint of this transition is $\approx$ 5~K, and we assign this temperature as $T_{\rm{LI}}$.

Fig.~\ref{Fig4} summarizes the temperature dependence of the magnetic Bragg peak at (0~0~4-$\tau$) compared with the bulk magnetization measurements of Ref.~~\onlinecite{Budko_2014}: (i) the feature observed at $\approx$ 7~K in the magnetization and specific heat measurements \cite{Budko_2014} correlates well with the values for $T_{\rm{N}}$ observed in the neutron diffraction data; (ii) the feature observed at $\approx$ 5~K in Fig.~\ref{Fig4}(a) is likely associated with $T_{\rm{LI}}$ and; (iii) the feature at $T^*$ $\approx$ 4~K, as we discuss below, is associated with the appearance of additional magnetic Bragg peaks with half-integer and integer indices identified, most clearly, in scans along the [0 1 $L$] and [0 2 $L$] directions.

First, to describe the magnetic structure associated with the (0~0~$\tau$) and (0~0~3$\tau$) magnetic Bragg peaks, we consider the elastic magnetic scattering cross-section which can be written as:

\begin{equation}\label{Eqn1}
    \frac{d\sigma}{d\Omega} = N\frac{(2\pi)^3}{V}(\frac{\gamma r_0 g}{2})^2 S(\overrightarrow{Q})
\end{equation}

and,

\begin{equation}\label{Eqn2}
    S(\overrightarrow{Q}) = |\sum_l f(\overrightarrow{Q}) [\widehat{Q} \times (\overrightarrow{\mu_l} \times \widehat{Q})]e^{-W_l}e^{i\overrightarrow{Q} \cdot \overrightarrow{R_l}}|^2
\end{equation}where the sum is over all magnetic ions in the magnetic unit cell, and $N$ and $V$ are the number and volume of the magnetic unit cells, respectively. The structure factor, $S(\overrightarrow{Q})$, depends upon the single ion magnetic form factor, $f(\overrightarrow{Q})$, the Debye-Waller temperature factor, $e^{-W_l}$, and the position, $\overrightarrow{R_l}$, and ordered moment, $\overrightarrow{\mu_l}$,  of the magnetic ions through $[\widehat{Q} \times (\overrightarrow{\mu_l} \times \widehat{Q})]e^{i\overrightarrow{Q} \cdot \overrightarrow{R_l}}$.  In particular, the component of the ordered moment directed parallel to the direction of momentum transfer, $\widehat{Q}$, does not contribute to the magnetic Bragg peak intensity, whereas the component directed perpendicular to $\widehat{Q}$ does, and this is often used as a means of determining the orientation of moments in magnetically ordered systems.  More generally, a complete evaluation of $S(\overrightarrow{Q})$, which takes both the structural twinning and magnetic domains into account, is required for a full solution of the magnetic structure.

\begin{figure}
\centering\includegraphics[width=0.9\linewidth]{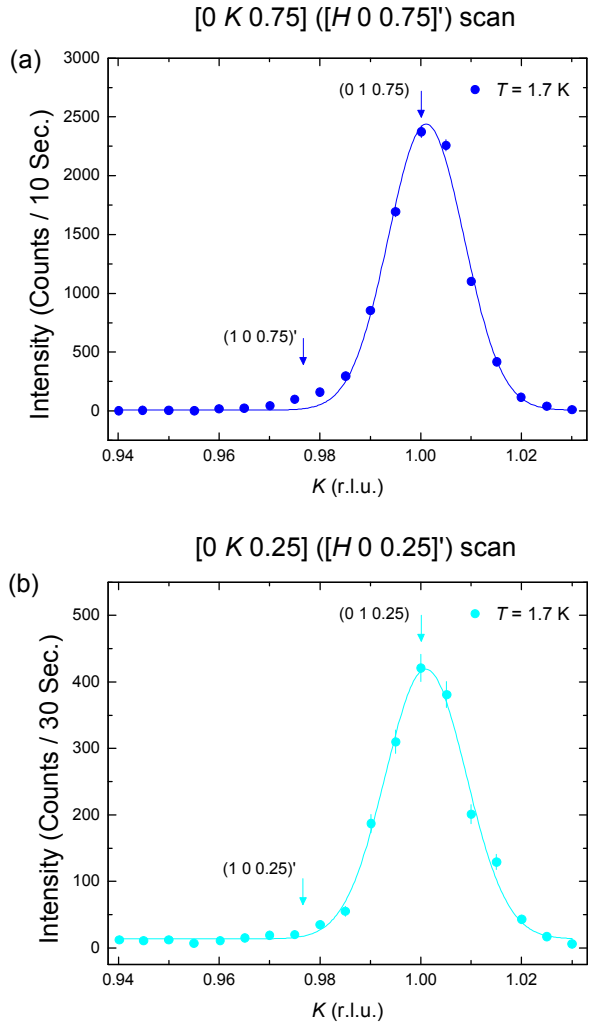}
\caption{(Color online) The $\tau$ and 3$\tau$ magnetic satellites at reciprocal lattice positions (a) (0~1~0.75) [(1~0~0.75)$^{\prime}$] and (b)(0~1~0.25) [(1~0~0.25)$^{\prime}$]. Scans were performed along the [0~$K$~0] ([$H$~0~0]$^{\prime}$) direction. Both magnetic peaks should appear because of the presence of both magnetic/structural domains. Magnetic scattering is observed at positions corresponding to the (0~1~0.75) and (0~1~0.25) Bragg peak, but absent (or very weak) at the (1~0~0.75)$^{\prime}$ and (1~0~0.25)$^{\prime}$ positions demonstrating that the ordered moment is along the \emph{\textbf{a}} axis of the orthorhombic structure.}
\label{Fig5}
\end{figure}

\begin{figure}
\centering\includegraphics[width=0.9\linewidth]{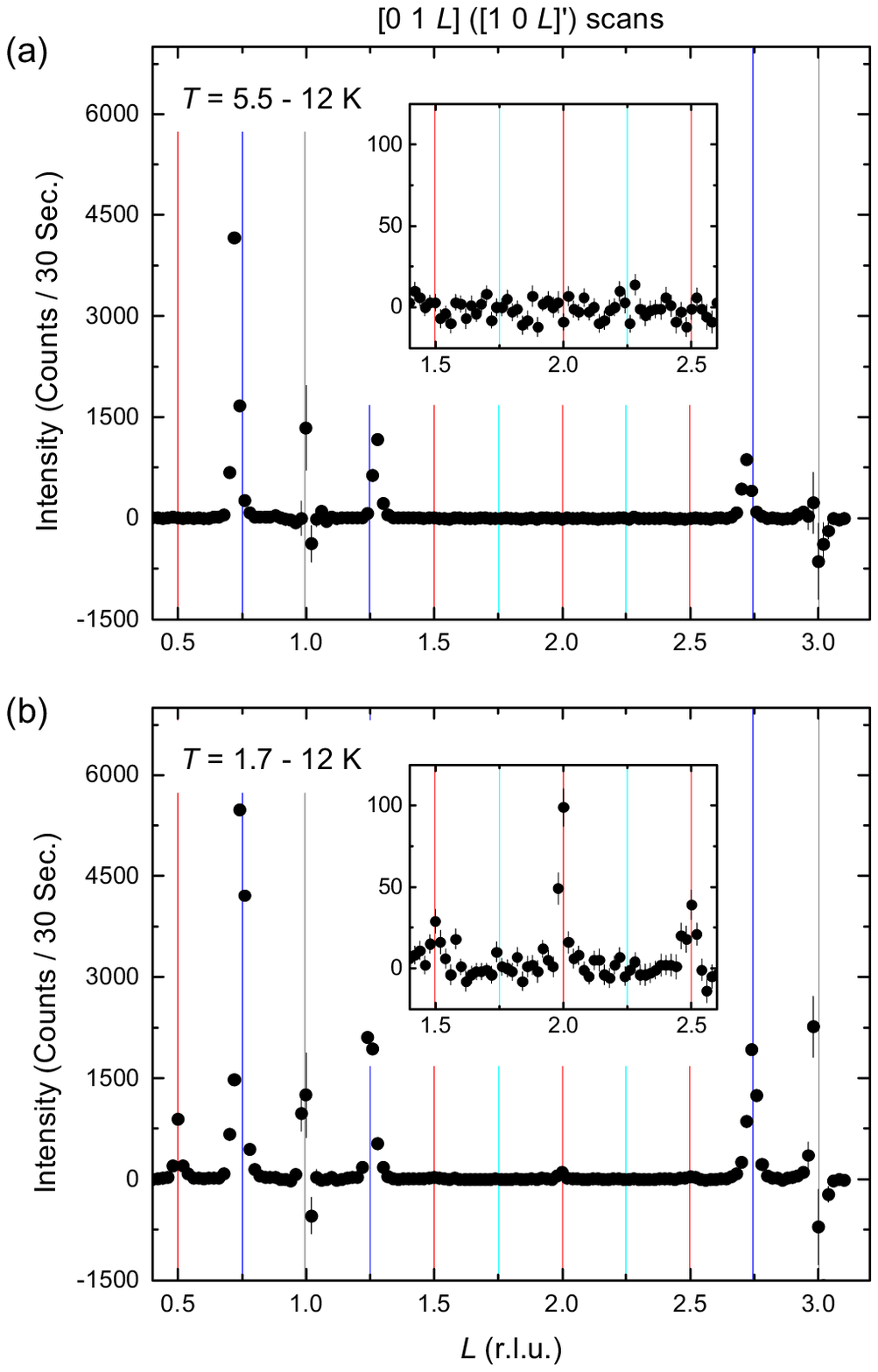}
\caption{(Color online) Scans along the [0~1~$L$] ([1~0~$L$]$^{\prime}$) reciprocal lattice direction showing (a) the difference between data taken at 5.5~K and 12~K and (b) the difference between data taken at 1.7~K and 12~K. The vertical lines denote the positions of half-integer and integer peaks in addition to the $\tau$ = $\frac{1}{4}$ and $\frac{3}{4}$ commensurate peak positions, as shown in the (0~$K$~$L$) reciprocal lattice plane map in Fig.~\ref{Fig2}(d).  The insets highlight the region near the half-integer and integer magnetic Bragg peaks.}
\label{Fig6}
\end{figure}

Figure~\ref{Fig5} shows scans through the (0~1~0.75) [(1~0~0.75)$^{\prime}$] and (0~1~0.25) [(1~0~0.25)$^{\prime}$] magnetic peak positions along the [0~$K$~0] ([$H$~0~0]$^{\prime}$) directions.  We remind the reader that both of these directions are probed in scans that are nominally along the [0~$K$~0] direction because of the presence of structural domains and twinning of the orthorhombic structure.  We see, however, that only the magnetic peaks indexed as (0~1~0.75) and (0~1~0.25) are observed whereas the (1~0~0.75)$^{\prime}$ and (1~0~0.25)$^{\prime}$ peaks, which should appear at a slightly lower value of $K$ because of the larger value of the $a$ lattice constant, are absent or very weak.  From Eqn.~\ref{Eqn2}, this most likely results from an orientation of the ordered magnetic moments along the \emph{\textbf{a}} axis of the orthorhombic unit cell.  Therefore, we can describe the magnetic structure as follows: For $T_{\rm{LI}}$~$<~T~<$~$T_{\rm{N}}$, the magnetic structure is incommensurate along the \emph{\textbf{c}} axis of the orthorhombic unit cell with a modulated moment along the \emph{\textbf{a}} axis.  As temperature is lowered, the incommensurate modulation squares up until, at $T_{\rm{LI}}$, the magnetic ordering locks in to a commensurate structure which can be described with a magnetic unit cell dimension along the \emph{\textbf{c}} axis which is four times that for the chemical unit cell.  The ordered moment remains along the \emph{\textbf{a}} axis.

\begin{figure}
\centering\includegraphics[width=0.9\linewidth]{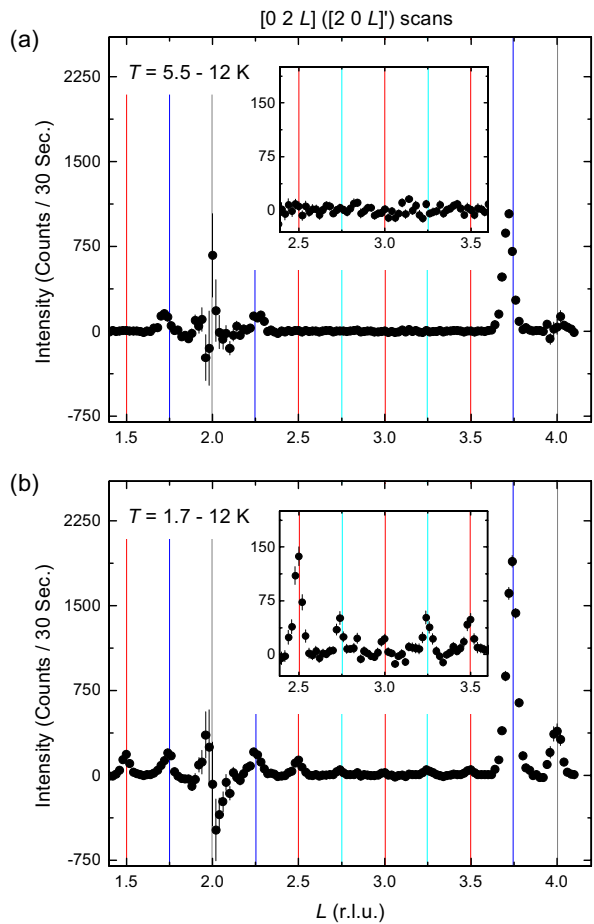}
\caption{(Color online) Scans along the [0~2~$L$] ([2~0~$L$]$^{\prime}$)  reciprocal lattice direction showing (a) the difference between data taken at 5.5~K and 12~K and (b) the difference between data taken at 1.7~K and 12~K. The vertical lines denote the positions of half-integer and integer in addition to the $\tau$ = $\frac{1}{4}$ and $\frac{3}{4}$ commensurate peak positions, as shown in the (0~$K$~$L$) reciprocal lattice plane map in Fig.~\ref{Fig2}(f).  The insets highlight the region near the half-integer and integer magnetic Bragg peaks.}
\label{Fig7}
\end{figure}

\begin{figure}
\centering\includegraphics[width=0.9\linewidth]{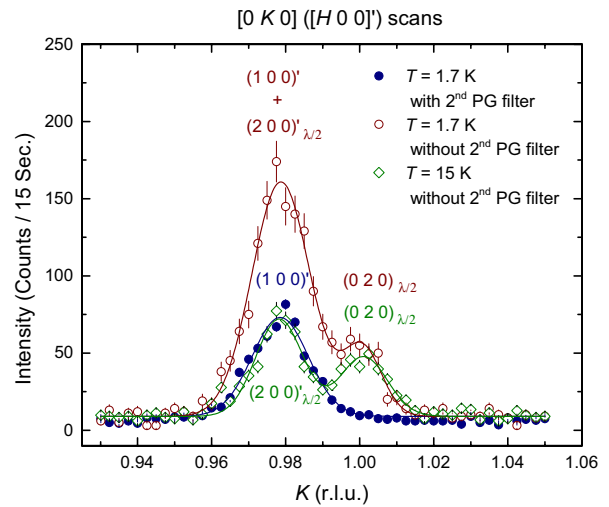}
\caption{(Color online) Scans along the [0~$K$~0] ([$H$~0~0]$^{\prime}$) directions through the (0~1~0) and (1~0~0)$^{\prime}$ peak positions. Both magnetic peaks should appear because of the presence of both magnetic/structural domains. Magnetic scattering is observed at the position corresponding to the (1~0~0)$^{\prime}$ Bragg peak, but not the (0~1~0) position demonstrating that the ordered moment is along the orthorhombic \emph{\textbf{b}}-axis.}
\label{Fig8}
\end{figure}

A series of scans along the [0~1~$L$] ([1~0~L]$^{\prime}$) and [0~2~$L$] ([2~0~L]$^{\prime}$) reciprocal lattice directions are shown in Figs.~\ref{Fig6} and \ref{Fig7} where we see, below $T^*$ $\approx$ 4~K, the appearance of half-integer [e.g. (0~1~2.5), (0~2~1.5)] and integer [e.g. (0~1~2), (0~2~3)] indexed diffraction peaks at positions forbidden for nuclear scattering.  The temperature dependence of the integrated intensity of the (0~1~0.5) reflection is also shown in Fig.~\ref{Fig4}. These new magnetic Bragg reflections are significantly weaker (on the order of a factor of 50) than those at the primary magnetic propagation vector (0~0~$\tau$) and are not clearly observed in the scans along the [0~0~$L$] direction [see the inset of Fig.~\ref{Fig2}(f)].

To further investigate the nature of the magnetic structure associated with these new magnetic diffraction peaks, scans were performed along the [0~$K$~0] ([$H$~0~0]$^{\prime}$) direction at the position of the (0~1~0) [(1~0~0)$^{\prime}$] magnetic Bragg peaks both with and without one of the PG filters.  Two PG filters effectively eliminate higher energy neutrons at $\frac{\lambda}{2}$ in the incident beam, so that only neutrons at the fundamental wavelength ($\lambda$) are detected.  Removing one of the filters allows some leakage of neutrons at $\frac{\lambda}{2}$ and, therefore, higher order nuclear Bragg peaks [e.g. (0~2~0) and (2~0~0)$^{\prime}$] can be seen at the (0~1~0) and (1~0~0)$^{\prime}$ magnetic Bragg peak positions. As shown in Fig.~\ref{Fig8}, scans taken without the PG filter at a temperature above $T_{\rm{N}}$ (open diamonds) show two Bragg peaks of comparable magnitude which arise from the (0~2~0) [(2~0~0)$^{\prime}$] reflections.  At $T$~=~1.7~K, (open circles) the intensity of the peak at slightly lower $K$ increases, whereas the intensity of the higher $K$ peak is unchanged.  Finally, with the second PG filter in place, eliminating the high energy harmonic content of the incident beam, we see that only the Bragg peak at (1~0~0)$^{\prime}$ is present (filled circles).

We assign the new half-integer and integer magnetic peaks to a second magnetically ordered state that coexists with the (0~0~$\tau$) magnetic state. First, the order parameter for these reflections is distinct from the dominant magnetic Bragg peaks associated with the (0~0~$\tau$) propagation vector (see Fig.~\ref{Fig4}).  Furthermore, based on the measurements described above, the moment associated with this new magnetic ordering lies along the orthorhombic \emph{\textbf{b}}-axis, perpendicular to the ordered moment determined for the Bragg peaks associated with the (0~0~$\tau$) magnetic propagation vector.  However, an unambiguous determination of this new magnetic structure below 4~K will require a full refinement of $S(\overrightarrow{Q})$ to determine why the half-integer and integer magnetic magnetic Bragg peaks are absent or very weak in the [0~0~$L$] scan below 1.9~K [see Fig.~\ref{Fig2}(f)].

As discussed in the Introduction, previous physical property measurements indicate the onset of FM order below 4~K in these samples [see Fig.~\ref{Fig4}(a)].  In our neutron data, this would be signalled by an increase in the intensity of the nuclear Bragg peaks below this temperature.  However, as can be seen in the difference plots of Figs.~\ref{Fig2}(e) and (f), as well as Figs.~\ref{Fig6} and \ref{Fig7}, no consistent positive difference could be resolved in our measurements.  This is not particularly surprising in light of the very small ferromagnetically ordered moment ($\approx$0.2 $\mu_{\rm{B}}$/Ce) estimated from the bulk magnetization measurements. To resolve this question, polarized neutron diffraction or x-ray magnetic circular dichroism measurements will be required.

Finally, scans along the diagonal directions in the ($H$~$K$~0) reciprocal lattice plane were performed to check for any reflections, indicative of long-range ordering of vacancies on the Ge sites (e.g. the $R_4$Ge$_7$ structure), at positions related to the (1/2~1/2~0) and (1/4~-1/4~0) chemical superlattice reflections described by Zhang \emph{et al.}\cite{Zhang_2013} for other $R$Ge$_{2-x}$ compounds.  No evidence of these superlattice peaks with an intensity in excess of 0.1\% of the strongest allowed nuclear reflections was found, consistent with previous studies of the CeGe$_{2-x}$ system.\cite{Zhang_2013} However, we can not exclude that short-range vacancy ordering may play some role in promoting the presence of the second AFM ordered state found in our measurements.  Regardless, we note that the AFM order reported here is related to propagation vectors along the orthorhombic \textbf{\emph{c}}-axis, whereas the reported chemical superstructures are related to propagation vectors of the chemical unit cell in the \textbf{\emph{ab}} plane.

\section {Summary}

We have performed neutron diffraction measurements on solution-grown single crystals of CeGe$_{1.76}$ between 300~K and 1.7~K to determine the nature of the magnetic ordering at low temperature.  Three characteristic temperatures have been identified at $T_{\rm{N}}$ $\approx$ 7~K, $T_{\rm{LI}}$ $\approx$ 5~K and $T^*$ $\approx$ 4~K corresponding to the transition from a paramagnetic phase above $T_{\rm{N}}$ to an incommensurate AFM structure characterized by a propagation vector (0~0~$\tau$) that persists down to $T_{\rm{LI}}$, where it locks-in to a commensurate value of $\tau$ = $\frac{1}{4}$.  The ordered moment was determined to be parallel to the orthorhombic \emph{\textbf{a}} axis.  Below $T^*$, a second magnetic state appears characterized by very weak magnetic Bragg peaks at half-integer and integer values along the [0~0~$L$] direction.  The data presented here indicate that the ordered moment in this magnetic state lie along the orthorhombic \textbf{\emph{b}}-axis, although further investigations are required for a full magnetic structure determination. All of the characteristic temperatures noted above are consistent with features observed in the bulk magnetization data of Ref.~~\onlinecite{Budko_2014}. However, we have found no consistent increase in the nuclear Bragg peak intensities below 4~K indicative of FM ordering although the proposed small ordered moment is likely below the present sensitivity of our measurements.

This research was supported by the U.S. Department of Energy, Office of Basic Energy Sciences, Division of Materials Sciences and Engineering. Ames Laboratory is operated for the U.S. Department of Energy by Iowa State University under Contract No.~DE-AC02-07CH11358. Research at ORNL's High Flux Isotope Reactor was sponsored by the Scientific User Facilities Division, Office of Basic Energy Sciences, U.S. Department of Energy.

\end{document}